\documentclass[11pt,twoside]{article}
\usepackage{asp2006}
\usepackage{epsf}
\usepackage{graphics}
\usepackage{lscape}
\def\edcomment#1{\iffalse\marginpar{\raggedright\sl#1\/}\else\relax\fi}
\reversemarginpar

\begin{document}
\title{The role of Starbursts in the Star Formation History of the Universe}
\author{Rosa M. Gonz\'alez Delgado}
\affil{Instituto de Astrof\'\i sica de Andaluc\'\i a (CSIC), E-18008 Granada,
Spain}

\begin{abstract}

This review presents a personal view of the role that starbursts play in the star formation history of the universe. It is mainly focused on the properties of nearby starburst galaxies selected for their strong UV and/or FIR emission. The similarities between local starbursts  and star-forming galaxies at high redshift are also presented.  I discuss too the role that LIRGs and ULIRGs  and merging systems play in the formation and evolution of galaxies.

\end{abstract}

\vspace{-0.5cm}
\section{Introduction}

\subsection{\bf Starburst  galaxies at high-redshift}

Starburst galaxies are systems with a high star formation rate, and the preferred place for the formation of massive stars.  Hence they are a relevant energy source that drives the cosmic evolution of galaxies. They are responsible of the thermal and mechanical heating of the interstellar medium, and its chemical enrichment. Because they have a significant large population of massive stars, they produce a large number of Lyman continuum photons that ionize the interstellar medium. But a fraction of these photons are absorbed by dust grain and re-emitted at the far-infrared wavelengths. 

Along the past  decade, the role that starbursts play in the star formation history of the universe has been highlighted mainly due to two important observational discoveries: a) a substantial population of star-forming galaxies at cosmological distances ($z\geq$ 2); b) the origin of the Cosmic Infrared Background (CIRB). These discoveries were made  thanks to the 
advance of the 10 m class telescopes, the launch of satellites sensitive at far-infrared wavelengths and the high spatial resolution and sensitive instruments on board $\it HST$.

Many of these high-redshift star-forming galaxies have been selected with the Lyman-break technique, that uses broad-band colors to target the Lyman continuum break of these galaxies (Steidel et al. 1996; Williams et al. 1996). This break has its origin in: a) the intrinsic drop in the spectral energy distribution of massive stars that dominate the UV light; b) the absorption by the neutral interstellar medium within the star-forming galaxy; c) the opacity of the intervening intergalactic medium. Subsequent spectroscopic observations with 10m class telescopes of these so called Lyman break galaxies (LBG) have shown that they are in a very active phase of star formation, and they contribute significantly to the global star formation rate (SFR) density of the universe (Madau et al. 1996). They are powered by massive stars, thus, hot O and B type stars dominate the integrated spectrum of these galaxies at ultraviolet (UV) wavelengths.
Their rest-UV spectra show a strong flat continuum which is plagued with low-ionization interstellar absorption lines, some high-ionization P Cygni wind lines, and a relatively weak Ly$\alpha$ emission that, in fact, is detected in absorption in many of these galaxies (Pettini et al. 2000; Shapley et al. 2003; Steidel et al. 2004; Noll et al. 2004). These spectral properties are very similar to those of local UV-bright starburst galaxies (Meurer et al. 1997; Gonz\'alez Delgado et al. 1998a; Heckman et al. 2005). 

Almost 25 years ago, the $\it IRAS$ satellite brought too a revolution in our understanding of the far-Infrared (FIR) emission of galaxies. A new class of galaxies was discovered that radiate the vast amount of their energy in the FIR. These galaxies, named luminous and ultraluminous far infrared galaxies (LIRGs and ULIRGs), have bolometric luminosities larger than  10$^{11}$ and 10$^{12}$ L$_\odot$, respectively. 
They are very rare objects in the local universe, and they contribute only with a few percent of the bolometric density in the local Universe. However,  more recent  observations obtained with $\it ISO$  show that bright infrared starbursts (LIRGs and ULIRGs) were more numerous in the past; being at z$\sim$ 1 the dominant contributor  of the most of the bulk of the CIRB emission (Elbaz \& Cesarsky 2003). Deep FIR surveys with $\it Spitzer$  have demontrated that the contribution of the LIRGs and ULIRGs to the total SFR density increases steadily with redshift
(Le Floc'h et al. 2005; P\'erez-Gonz\'alez et al. 2005), being LIRGs and ULIRGs  the most significant contributors to the infrared galaxy population at z$\geq$1. These galaxies appear to be the high-z analogs to the LIRGs and ULIRGs in the local Universe. 

These similarities between high redshift and local starbursts highlight the cosmological relevance of starburst galaxies. However, the importance of studying starburst galaxies goes beyond this. Nearby starbursts are the laboratories where to study violent star formation processes and their interaction with the interstellar and intergalactic medium.

\subsection{\bf Starburst galaxies: Definition and general properties}

Local  starbursts are mainly selected on the basis of their nebular optical emission lines, strong UV continuum, and/or  
far-infrared radiation. These criteria imply that the starburst class is formed by a mixed type of star-forming systems that include galaxies with starburst event and real starburst galaxies  (e.g. HII regions like 30Dor, blue compact dwarf galaxies, nuclear starbursts, and LIRGs and ULIRGs).  

Recently, Heckman (2005) has proposed that the most useful parameter for defining a  starburst is its intensity (SFR/area). This definition allows us to compare local starbursts with physically similar galaxies at high-z. So, starburst galaxies must have SFR per unit area which is  larger than that in the disks of normal galaxies.  Nearby starbursts have  star formation intensities, ranging from 1 to 100 M$_\odot$ yr$^{-1}$ kpc$^{-2}$, and similar values are found for LBGs (Meurer et al. 1997). Recently,  observations with $\it GALEX$ have provided evidence of the existence of true local starbursts (0.1$\leq$ z $\leq$ 0.3) which are the analogs to the LBGs in terms of their size and UV luminosity.  They are compact ultraviolet luminous galaxies (I$\rm_{FUV}\geq 10^8$ L$_\odot$ kpc$^{-2}$), low-mass galaxies (M$_{star}$ $\sim$ 10$^{10}$ M$_\odot$) with half-light radii less than a few kpc. They have large enough star formation rates to build the present galaxy in $\sim$1--2 Gyr. They also have metallicity, velocity dispersion and extinction similar to the LBG population (Heckman et al. 2005).

\section{UV selected starbursts}

Observations of starbursts at UV wavelengths are very relevant because they allow us a direct detection of massive stars, and thus a direct measurement of the star formation rate. This spectral range is also quite sentitive to the star formation history of galaxies, and the evolution of stellar populations up to intermediate ages ($\sim$ 1 Gyr).
Dust, however,  has a profound effect on the UV emission in starbursts. $\it {GALEX}$ observations of a sample of UV selected local galaxies have shown that only 33$\%$ of the UV emission escapes from the galaxies and the remaining 66$\%$ is absorbed by dust and reradiated in the far-infrared (Buat et al. 2005). But, due to the 'picket-fence' interstellar dust distribution, starbursts can be bright at ultraviolet (UV) wavelengths, as they are in the far-infrared. Two good examples of nearby starbursts  that are strong UV and FIR (LIRGs) galaxies are IRAS0833+65 and NGC 6090 (see Figure 1).   

\begin{figure}[htbp]
%\special{psfile=f1_gonzalezdelgado.eps angle=-90 vscale=55 hscale=55 hoffset=-30 voffset=30 }
\includegraphics{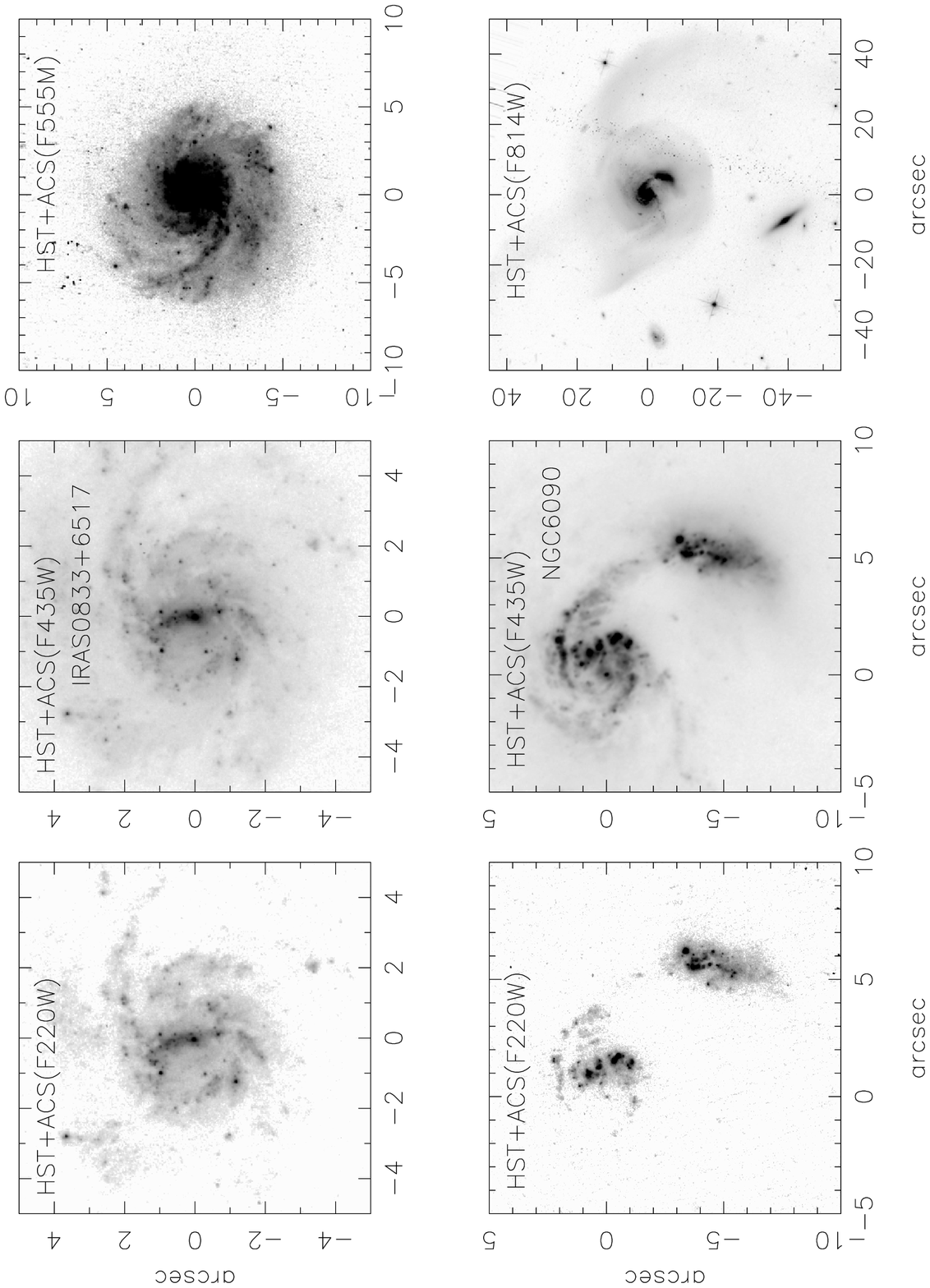}
\vspace{10cm}
\caption{$\it HST+ACS$ images taken through F220W (left), F435W (center) and F555W or F814W (right) filters for two bright UV selected starbursts. Stellar cluster knots and diffuse emission are clearly detected at UV and B bands. }
\label{fig:f1}
\end{figure}                        

In this section the UV continuum and L$\alpha$ properties of nearby starbursts are discussed. A more extended review about starbursts at UV wavelengths is in Gonz\'alez Delgado (2006).

\subsection{\bf UV continuum properties}

 At UV wavelengths, many starbursts show a continuum filled with absorption lines including resonant transitions of many ionic species. These features have three different origins: stellar winds, stellar photospheres, and the interstellar medium. In contrast, at optical wavelengths the spectrum can be dominated by nebular emission lines;  although photospheric lines, such us high-order Balmer lines, mainly produced by B and A stars can be detected also at the optical.
 
 Massive stars develop strong wind stellar lines due to the radiation pressure in ultraviolet resonance lines. As a result, all the strong ultraviolet lines show a blueshifted absorption (about 2000--3000 km s$^{-1}$) or a P Cygni profile. The shape of the profile reflects the stellar mass-loss rate, which is related to the stellar luminosity, and thus to the stellar mass content of the starburst.  The most relevant wind lines in starbursts are: NV $\lambda$1240, SiIV $\lambda$1400, CIV $\lambda$1550, and HeII $\lambda$1640.  Photospheric lines form in layers that are in hydrostatic equilibrium. Mainly from C, N, O, S, Si and Fe ions of low and high-excitation potential, they originate from excited levels, mainly from C, N, O, S, Si and Fe ions of low and high-excitation potential. They are weaker than the wind lines and depend on the stellar population age and metallicity.  Some of the most relevant of these lines in starbursts are: SV $\lambda$1502, CIII $\lambda$1426-1428 and CIII $\lambda$1176.  Wind lines are very useful diagnostics to constrain the stellar population properties (IMF, and age) in starbursts.  Wind and photospheric lines are very useful diagnostic to constrain the stellar population properties (IMF, and age) in starbursts. HST observations have revealed that a Salpeter IMF with high-mass stars constrain well the UV properties of starbursts, at low and high-metallicities.
 
 Most of the low ionization resonant lines are of interstellar origin. These lines are very useful to study the interaction of the massive stars with the interstellar medium, and they often reflect the hydrodynamical impact of the starbursts on the interstellar medium. Large scale outflows produced by the starburst have left their imprints on the UV interstellar lines. Outflows of a few hundred km s$^{-1}$ are ubiquitous in nearby starbursts and in LBG (e.g Gonz\'alez Delgado et al. 1998; Shapley et al. 2003).
 
HST has been the first telescope to provide UV high spatial resolution images of nearby starbursts. These images show that the UV continuum in starbursts is provided by two components, compact knots and diffuse emission. The compact knots  provide about 25$\%$ of the total UV emission (Meurer et al. 1995). They have magnitudes and sizes that suggest that they are stellar clusters. These clusters are distributed irregularly over the diffuse background that extends  for about a few 100 pc (Figure 1). Several origins have been proposed for the UV diffuse emission: a) Continuous star formation lasting for a few 100 Myr; originally, the stars form in clusters over the last few 100 Myr, but clusters dissolve with age and disperse across the field. b) The UV radiation originates in dusty compact stellar clusters, but it is scattered  by dust to the field. c) Individual massive stars, unresolved even at the HST spatial resolution. Long-slit UV spectra of starbursts taken with STIS point to the star formation origin, so that the UV diffuse radiation is created via dissipation of aging star clusters (Tremonti et al. 2001; Chandar et al. 2005). These results are very relevant because indicate: a) stellar clusters are the most important mode of star formation in starbursts; b) intermediate age stars (B and A), that result from  older dissipated clusters, contribute at the optical continuum (as indicated by  the detection of high-order Balmer lines (HOBL) in absorption, see Figure 4) in  many of these nearby starbursts; c) the UV continuum is a good SFR indicator in these objects, because the UV emission is not  radiation scattered by the clusters.

\subsection{\bf The Ly$\alpha$ line: outflows of neutral H gas}
 
 The massive stars in the starbursts provide enough ionizing photons to produce the Ly$\alpha$ line in emission
 by recombination of hydrogen photoionized.  But  Ly$\alpha$ photons are attenuated by dust as a result of multiple resonant scatterings by hydrogen atoms, that increase the path length of the Ly$\alpha$ photons and thus the probability that they will be absorbed by dust. Extinction also affects the line strength. However, in most of the nearby starbursts,  Ly$\alpha$ is detected with a broad damped absorption or a P-Cyg profile. These results suggest that dust and/or metallicity are not the key parameters for determining the strength and shape of  Ly$\alpha$. UV $\it HST$ spectra have shown that the line is asymmetric, with the peak of the emission redshifted with respect to the systemic velocity, and a deep absorption which is blueshifted by several 100 km s$^{-1}$ with respect to the emission (Gonz\'alez Delgado et al. 1998; Mas-Hesse et al. 2003). This absorption shift is in agreement with the blueshift observed in the interstellar lines in these galaxy. The natural explanation is that the neutral HI gas producing the absorption is outflowing from the starburst. Hydrodynamic calculations have proved that the properties of the Ly$\alpha$ line are linked to the evolutionary state of the starburst (Tenorio-Tagle et al. 1998). 
 
The Ly$\alpha$ properties of LBGs are very similar to  nearby starbursts. Shapley et al. (2003) divided a sample of $\sim$ 1000 LBGs in four classes according with the equivalent width of the line, Ew(Ly$\alpha$). They find that LBGs with Ew(Ly$\alpha$) $\geq$ 20 \AA\ have bluer UV continuum, weaker low-ionization interstellar absorption lines, smaller kinematic offset between Ly$\alpha$ and the interstellar lines, and lower star formation rates than LBGs as a whole. They suggest that the appearance of LBGs spectra is determined by a combination of the covering fraction of outflowing  neutral gas, which contains dust, and the range of velocities over which this gas is absorbing.

In summary, like the interstellar lines, Ly$\alpha$ line in nearby starbursts and LBGs is driven by the dynamical effects of the violent star formation processes ongoing in the starburst, rather than by the gravitational potential of the galaxy.

 \section{FIR selected starbursts}
 
\subsection{Nuclear activity, starbursts and mergers}

 One key point  of researching in LIRGs and ULIRGs since they were discovered is the origin of the IR emission. Optical spectra of these galaxies show emission line ratios that indicate that they host a starburst and/or an AGN. Moreover, it has been found that the fraction of galaxies with signs of Seyfert activity increases significantly with the FIR luminosity. So, LIRGs are mainly starbursts, but about half of the ULIRGs have a Seyfert nucleus (e.g. Veilleux, Kim, and Sanders 1999). However, mid-infrared spectroscopy studies indicate that most of the cool ULIRGs ($f_{\rm 25}/f_{\rm 60}\leq$ 0.2) are powered by a starburst, even they host an AGN (e.g. Farrah et al. 2007). The origin of  the power source in warm ULIRGs (most of them are QSOs),  is still controversial.  
 
 Other important research issue in LIRGs and ULIRGs is  related with the role that these objects play in the formation and evolution of galaxies. In the context of the hierarchical galaxy formation and evolution models, mergers can trigger the star formation and the nuclear activity in galaxies, and lead to the formation of elliptical galaxies. So, as galaxies merge to form larger units, gas is driven into the nuclear region triggering a starburst and growing a supermassive black hole (BH). In this scenario the creation and evolution of a BH (presumably associated to a QSO) is intimately connected to that of the galaxy bulge. Evidences in favor of this scenario are the tight correlation between the BH mass and the bulge stellar velocity dispersion (Ferrarese and Merrit 2000) and the similarities between the evolution of the QSO luminosity density and the star formation rate (Boyle and Terlevich 1998).
 
 At the local universe, there is significant evidence of the link between ULIRGs  and galaxy interactions and mergers. Veilleux et al. (2002) have found that ULIRGs are mergers of gas-rich disk galaxies  in the later stages. Moreover, Borne et al (2000) found that ULIRGs can be at different merger stages since many of them are in dense galaxy groups. LIRGs morphology is, however, more diverse (Arribas et al. 2004). Recent $\it HST$ observations with the ACS show that many LIRGs, as  ULIRGs, mergers in the later stage (e.g. NGC 6240), but some of them show very little signs of interactions (e.g. NGC 6436). Additionally, LIRG morphologies encompass all the expected merging stages (Figure 2). At intermediate redshift, half of the LIRGs show a spiral disk morphology and only one third of these galaxies are mergers (Elbaz et al. 2005;  Hammer et al. 2005).
 
 \begin{figure}[htbp]
%\special{psfile=f2_gonzalezdelgado.eps angle=-90 vscale=55 hscale=55 hoffset=-30 voffset=30 }
\includegraphics{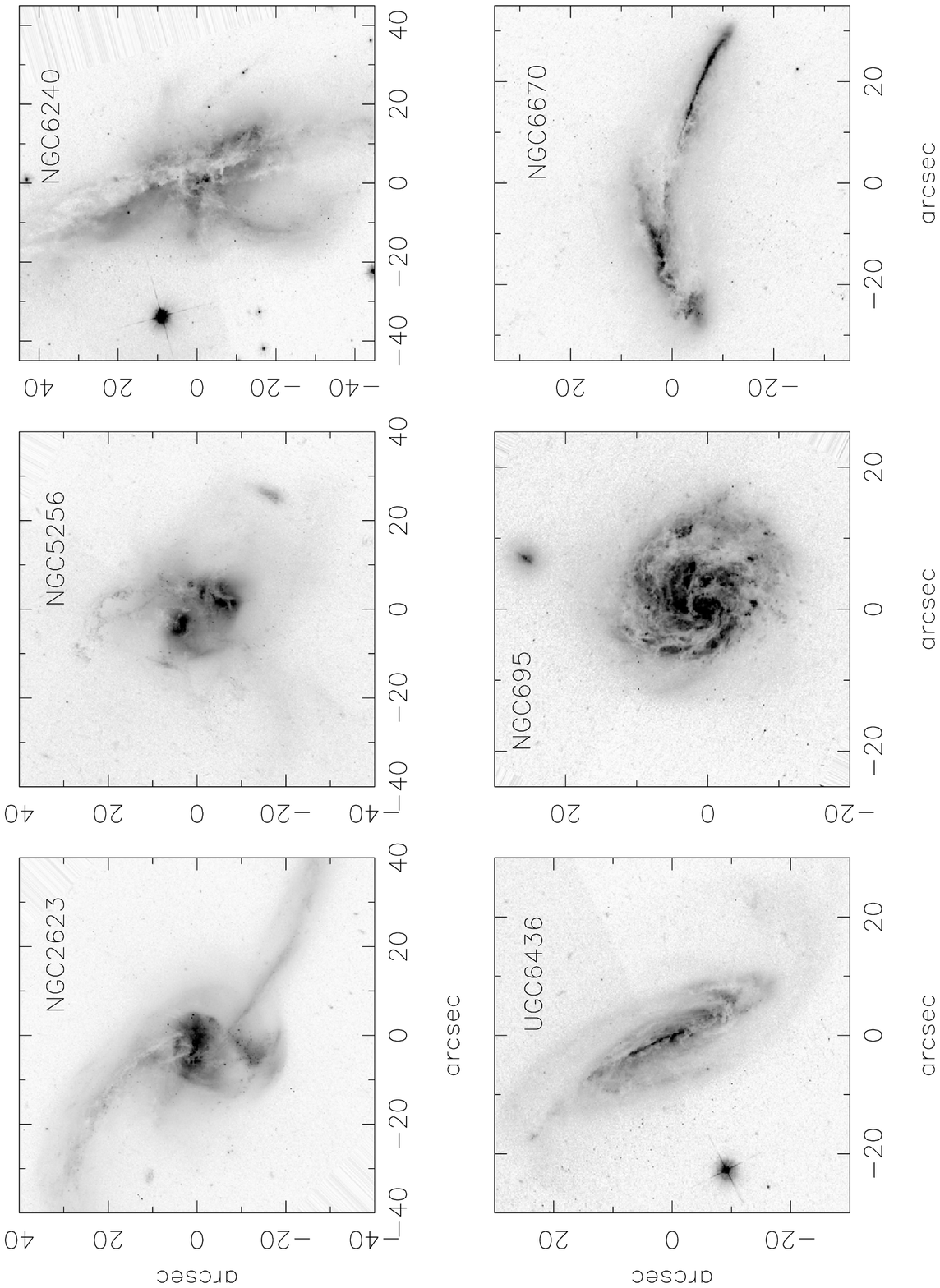}
\vspace{19cm}
\caption{$\it HST+ACS$ images taken through F435W of LIRGs. Examples of LIRGs that are disk, interacting or merger systems are shown. }
\label{fig:f1}
\end{figure} 
 
 \subsection{Evolutionary link between ULIRGs, Radio Galaxies, QSOs and elliptical galaxies}
 
 Theoretical models also predict that major disk-disk mergers can form ellipticals (e.g. Toomre and Toomre 1972). Evidence in favor of this scenario come from the HI and optical morphology and kinematics, and stellar population properties  observed in some proto-ellipticals galaxies (e.g. Schweizer 1996). The optical spectra show stellar lines typical of post-starburst galaxies. These features indicate that a major starburst occured about 1 Gyr. 
 
Kinematic observations of ULIRG mergers have proved that  they are ellipticals in formation. Measurements of their stellar velocity dispersion, surface brightness and effective radius indicate that ULIRGs follow well the fundamental plane, and they can form intermediate mass (10$^{11}$ M$_\odot$) elliptical galaxies (Genzel et al. 2001). In this scenario, mergers can drive an evolutionary sequence from cool ULIRGs to warm ULIRGs in their way to becoming quasars. Thus, mergers go first through a starburst phase, followed by a dust enshrouded Seyfert phase, then the dust is cleaned by the AGN and starburst wind actions, and finally ULIRGs are shown as QSOs (Sanders et al. 1988). 

There is also evidence that radio galaxies can be part of the evolutionary sequence between ULIRGs and elliptical galaxies. 
Even though most of the radio galaxies are ellipticals or S0 types, optical imaging of radio galaxies show double nuclei, tidal tails, arcs and distorted isophots in about half of the powerful FRII radio galaxies (Heckman et al. 1986). Kinematic evidences also support this hyphotesis (e.g. Tadhunter, Fosbury and Quinn 1989). 

Brotherton el al. (1999) have identified a class of QSOs, named post-starburst QSOs, most spectacular whose is UN J1025-0040 (z=0.634). Its representative spectrum is a chimera because it displays the broad MgII $\lambda$2800 emission line and a strong continuum characteristic of classical QSOs. But the optical continuum displays a large Balmer break, prominent high-order Balmer absorption lines, and CaII K and H lines. These stellar features and the bolometric luminosity are consistent with the object having been a ULIRG a few 100 Myr ago. More cases with similar optical spectra have been identified in the SDSS data releases.  These are key objects  to prove  the evolutionary link between ULIRGs and QSOs.  HST images (Figure 3) reveal that the QSO host resembles a merger remnant. Thus, these objects could be in a transition phase between ULIRGs and classical QSOs. 

 \begin{figure}[htbp]
\includegraphics{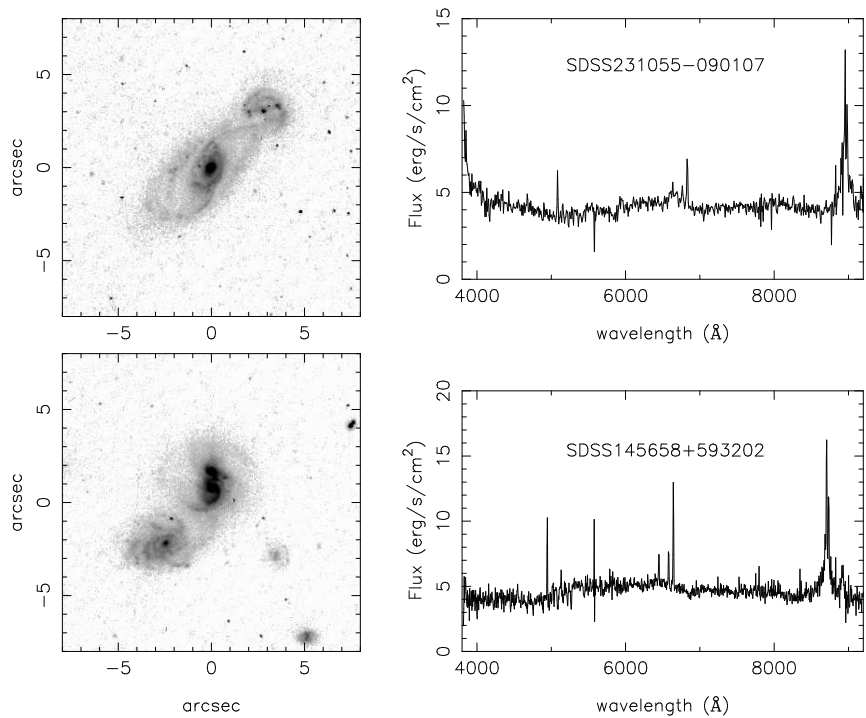}
\vspace{9.5cm}
\caption{$\it HST+ACS$ images and SDSS spectra of two post-starburst QSOs. }
\label{fig:f1}
\end{figure}

  \subsection{Stellar Populations in VLIRGs, ULIRGs, and Radio Galaxies}
  
To find out  an evolutionary link between  these objects  we are performing an optical spectroscopic program to 
 investigate the evolution of the stellar populations in several samples of VLIRGs, ULIRGs, radio galaxies and post-starburst QSOs. We use the age of the stellar populations as a clock to set a possible evolutionary sequence between them. This technique has been used previously to find out the possible evolution from Seyferts and low-luminosity AGNs (e.g. Gonz\'alez Delgado et al. 2001, 2004; Cid Fernandes et al. 2005a) and from ULIRGs-QSOs to classical QSOs (Canalizo and Stockton 2001). 
 
 The optical spectral range is rich in photospheric stellar lines that are very useful  to constrain the young and intermediate age population properties.  In particular, HOBL, and the Balmer break are very useful diagnostics  to date starburst and post-starbursts (Gonz\'alez Delgado, Leitherer and Heckman 1999).  However, the fit of the whole optical range (including continuum shape and stellar lines) instead of only the equivalent widths of a few  stellar lines is the best method to constrain the stellar population properties of galaxies. It has been clearly demonstrated recently using SDSS data of several hundred thousands of local early type and star-forming galaxies by Cid Fernandes et al. (2005b). This method applies well to the objects discussed here because the extinction can modify drastically their continuum shape; and moreover, the strength of some of  stellar lines are affected by nebular contribution. 
 
 Evolutionary synthesis models have proved to be a very useful tool to constrain the stellar population properties in galaxies. Models  that incorporate intermediate (e.g. Bruzual and Charlot 2003) or high spectral resolution stellar libraries (e.g.  Gonz\'alez Delgado et al. 2005) are really required to be able to isolate the young, intermediate and old population contributions, and to distinguish between age and extinction effects. 
 
 Other key issue to determine the stellar population properties is the technique used to find the best model that fits the whole observed spectrum. There are a staggering variety of techniques, and codes developed with this purpose. Here we use 
 STARLIGHT (Cid Fernandes et al. 2005b) that has proved to give very reasonable results (Cid Fernandes et al. 2005a). The STARLIGHT technique consists in fitting the observed spectrum with a combination of several simple stellar populations (SSP) from evolutionary synthesis models. Extinction is modeled by a foreground dust, and parametrized by the V-band extinction A$_{\rm V}$. The output is the population vector that represents the fractional contribution of each SSP of a given age and metallicity to the model flux at a given wavelength. The fit is carried out with a simulated annealing plus Metropolis scheme, which searchs for the minimum chi-square between the observations and the combined models.
  
Here, in this work, we have adopted for each metallicity a base of 14 SSPs with ages 4, 5, 10, 25, 40, 100, 280, 500, 900 Myr, and 1.26 and 14 Gyr from the models of Gonz\'alez Delgado et al. (2005). The Calzetti et al.  (2000) law was adopted. We also include several additional stellar components that correspond to SSP of ages $\leq$ 10 Myr and are reddened by one, two or three magnitudes. The purpose of this is  to allow the code to fit young SSPs with larger extinction than the intermediate and old SSPs. 
 
The results obtained for VLIRGs and ULIRGs are summaried as: a) the old stellar populations (age $\geq$ few Gyr)  contributes very little to the total optical continuum; b) very young ($\leq$ 10 Myr) plus intermediate (100-900 Myr) age stellar population can account for almost all the optical light; c) extinction is significant for these objects, and the young stellar population is redder than the intermediate age. 
As expected, the extinction is high. It is noted because: a) the continuum is very red even though the stellar features, such as HOBL are typical of young and/or intermediate age populations; b) the NaI line is very deep, indicating that its origin is mainly interstellar. It means that the source is partially covered by large column densities of the interstellar gas, and in consequence, the extinction is high. Results obtained for several VLIRGs and the ULIRG Arp 220 are presented in Figure 5. 
 
 We have found, like in VLIRGs and ULIRGs, strong evidence for intermediate (0.3-2.5 Gyr) and young ($\leq$ 100 Myr) age stellar populations in the nuclear region of radio galaxies (Tadhunter et al. 2005; Holt et al. 2007). These stellar populations are reddened but significantly less than VLIRGs and ULIRGs. These populations are very massive, and they could represent  starbursts that were able to heat the dust and produce a fraction of the far-infrared luminosity  detected with $\it Spitzer$  for these objects (Tadhunter et a. 2007).
 
 We have now taken 3D optical data with VIMOS at the VLT and PMAS at CAHA observatory to determine the host stellar population properties of the post-starburst QSOs.  The age is the key parameter to find out whether an evolutionary link 
exists between ULIRGs, and post-starburst QSOs. This is a work in progress, and the results will be reported  soon.

\begin{figure}[htbp]
\includegraphics{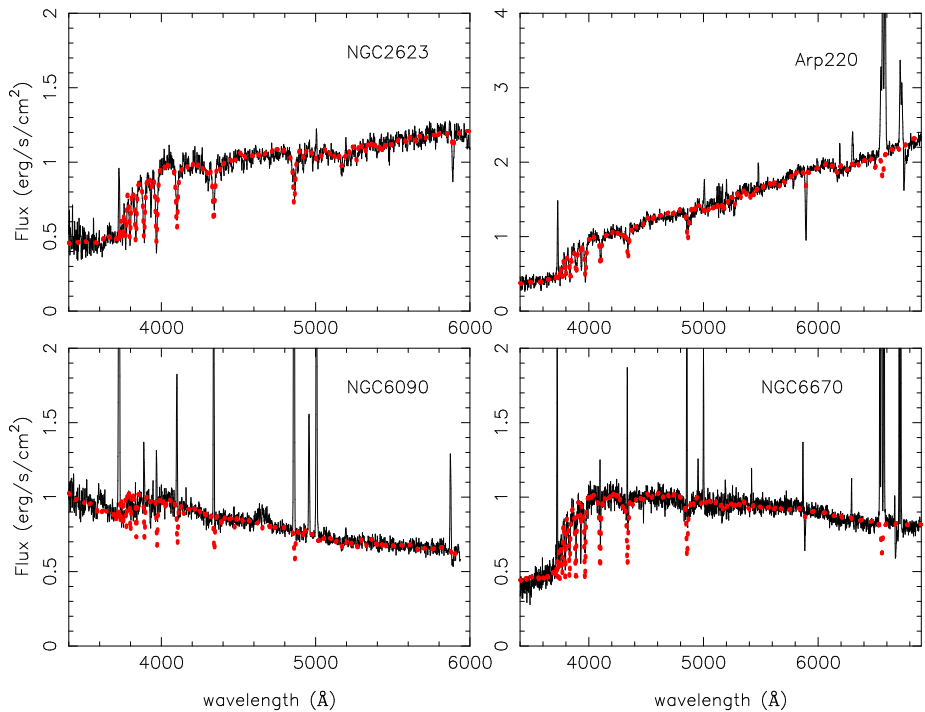}
\vspace{9.5 cm}
\caption{Optical spectra of some LIRGs (NGC6090, NGC6670, NGC2623) and the archetitypical ULIRG Arp220. The best models found with STARLIGHT and the high-spectral resolution evolutionary synthesis models from Gonz\'alez Delgado et al (2005) are plotted as dotted lines.}
\label{fig:f1}
\end{figure} 

 \section{Summary and Future prospects}
 
This review provides evidence of the role that  starbursts play in the star formation history of the universe, and  the origin and evolution of galaxies. In this decade, many high redshift galaxies that are in a starbursting phase have been discovered. This highlights the cosmological relevance of starbursts. But it is still relevant to observe nearby starbursts because they 
are the preferred laboratories where to study violent star formation processes and their interaction with the interstellar medium.  Near and far-starbursts can be triggered by mergers and interactions between galaxies; VLIRGs and ULIRGs are good examples of these processes. The stellar population properties of VLIRGs and ULIRGs indicate that they can be elliptical galaxies in formation. 

The future prospects are bright. The Spanish community can contribute significantly to this field thanks to the inminent operational stage of the GTC (official inauguration is in July 2007) and the extensive cosmological survey, ALHAMBRA (Moles et al. in these proceedings), that we are doing at the 3.5m CAHA observatory. ALHAMBRA will provide accurate redshifts for 6.6$\times$10$^5$ galaxies with $I\leq$ 25 and mean redshift of 0.74. Thanks to the 20 contiguous medium band optical filters observations many emission lines galaxies will be detected; and their spectral energy distribution will be used to trace the star formation history of the universe from the local to intermediate redshift. 

OSIRIS (Cepa et al. 2005) at the GTC will be a crucial instrument to probe the feedback processes taking place in  starbursts and AGNs. Deep imaging taken with OSIRIS tuned at the emission lines Ly$\alpha$, H$\alpha$, and optical lines (such as [OIII], [NII], [SII]) of Starbursts (ULIRGs-VLIRGs) and QSOs will be crucial to determine whether there is an evolutionary link between star-formation and nuclear activity.

 \subsubsection{Acknowledgements}
 
 I am very grateful to John and Johan for inviting me to this lovely conference. 
 I would like to thank my collaborators Roberto Cid Fernandes, Enrique P\'erez, Javier Rodr\'\i guez, Clive Tadhunter and Joanna Holt. This work is supported by Spanish grant AYA2004-02703.


\begin{thebibliography}{}
  
\bibitem[]{} Arribas, S., Bushouse, H., Lucas, R.A., Colina, L., Borne, K.D.  2004, AJ, 127, 2522

\bibitem[]{} Borne, K.D., Bushouse, H., Lucas, R.A., Colina, L. 2000, ApJ, 529, 77

\bibitem[]{} Boyle, B.J. \& Terlevich, R.J. 1998, MNRAS, 293, L49

\bibitem[]{} Brotherton, M.S., van Breugel, W.,  Stanford, S.A., Smith, R.J., Boyle, B.J., Miller, L.,  Shanks, T., Croom, S.M.,  Filippenko, A.V. 1999, ApJ, 520, L87

\bibitem[]{} Bruzual, G. \& Charlot, S. 2003, MNRAS, 344, 1000
 
\bibitem[]{}  Buat,  V., Iglesias-P\'aramo, J., Seibert, M.  et al. 2005, ApJ, 619, L63

\bibitem[]{} Calzetti, D., Armus, L.,  Bohlin, R.C., Kinney, A.L., Koornneef, J., Storchi-Bergmann, T. 2000, ApJ, 533, 682

\bibitem[]{} Canalizo, G. \& Stockton, A. 2001, ApJ, 555, 719

%\bibitem[]{} Cardelli, J.A., Clayton, G.C.,  \& Mathis, J.S. 1989, ApJ, 345, 245

\bibitem[]{} Cepa, J.; Aguiar, M.; Casta–eda, H. O. et al. 2005, RMxAC, 24, 1

\bibitem[]{} Chandar, R., Leitherer, C.,  Tremonti, C.A., Calzetti, D., Aloisi, A.,  Meurer, G.R., de Mello, D. 2005, ApJ, 628, 210

\bibitem[]{} Cid Fernandes, R., Gonz\'alez Delgado, R. M.,  Storchi-Bergmann, T., Martins, L. Pires,  Schmitt, H. 2005a, MNRAS, 356, 270

\bibitem[]{} Cid Fernandes, R. Mateus, A.,  Sodr\'e, L.,  Stasi\`nska, G.,  Gomes, J.M. 2005b, MNRAS, 358, 363
 
\bibitem[]{} Elbaz, D. \& Cesarsky, C.J.  2003, Sci, 300, 270

\bibitem[]{} Elbaz, D., Le Floc'h, E., Dole, H.,  Marcillac, D. 2005, A\&A, 434, L1

%\bibitem[]{} Farrah, D., Bernard-Salas, J., Spoon, H.W.W.  et al. 2007, ApJ, in press (astro-ph/07060513)
\bibitem[]{} Farrah, D.  et al. 2007, ApJ, in press (astro-ph/07060513)

\bibitem[]{} Ferrarese, L. \& Merritt, D. 2000, 539, L9

\bibitem[]{} Genzel, R., Tacconi, L.J., Rigopoulou, D., Lutz, D., Tecza, M. 2001, ApJ, 563, 527

\bibitem[]{} Gonz\'alez Delgado, R.M., 2006, Ap\&SS, 303, 58
  
\bibitem[]{} Gonz\'alez Delgado, R.M., Leitherer, C., Heckman, T., Lowenthal, J.D., Ferguson, H.C. Robert, C. 1998, ApJ, 459, 698

\bibitem[]{} Gonz\'alez Delgado, R.M., Leitherer, C. \& Heckman, T. 1999, ApJS, 125, 489

\bibitem[]{} Gonz\'alez Delgado, R.M., Heckman, T., Leitherer, C. 2001, ApJ, 546, 845

\bibitem[]{} Gonz\'alez Delgado, R.M.,  Cid Fernandes, R., P\'erez, E., Martins, L.P., Storchi-Bergmann, T., Schmitt, H., Heckman, T., Leitherer, C. 2004, ApJ, 605, 127

\bibitem[]{} Gonz\'alez Delgado, R.M., Cervi\~no, M., Martins, L.P., Leitherer, C., Hauschildt, P.H. 2005, MNRAS, 357, 945

\bibitem[]{} Hammer, F., Flores, H., Elbaz, D.,  Zheng, X.Z.,  Liang, Y.C.,  Cesarsky, C. 2005, A\&A, 430, 115

\bibitem[]{} Heckman, T.M., Smith, E.P., Baum, S.A., van Breugel, W.J.M., Miley, G.K., Illingworth, G.D., Bothun, G.D., Balick, B. 1986, ApJ, 311, 526

\bibitem[]{} Heckman, T.M.  2005, Starbursts: From 30 Doradus to Lyman Break Galaxies. Astrophysics \& Space Science Library, 329, 3, (eds. de Grijs, R., Gonz\'alez Delgado, R.M.)

\bibitem[]{} Heckman, T., Hoopes, C.G., Seibert, M., et al. 2005, ApJ, 619, L35
 
\bibitem[]{} Le Floc\'h, E.,  Papovich, C.,  Dole, H., et al. 2005, ApJ, 632, 169 
 
\bibitem[]{} Madau, P., Ferguson, H.C., Dickinson, M.E., Giavalisco, M., Steidel, C.C., Fruchter, A. 1996, MNRAS, 283, 1388

\bibitem[]{} Mas-Hesse, J.M., Kunth, D., Tenorio-Tagle, G., Leitherer, C., Terlevich, R.J., Terlevich, E. 2003, AJ, 598, 858
 
\bibitem[]{}  Meurer, G.R., Heckman, T.M., Leitherer, C., Kinney, A., Robert, C., Garnett, D.R. 1995, AJ, 110, 2665
 
\bibitem[]{} Meurer, G.R., Heckman, T.M., Lehnert, M.D.; Leitherer, C., Lowenthal, J. 1997, AJ, 114, 54
 
\bibitem[]{} Noll, S., Mehlert, D., Appenzeller, I., et al. 2004, A\&A, 418, 885
  
%\bibitem[]{} P\'erez-Gonz\'alez, P.G., Rieke, G.H., Egami, E. et al. 2005, ApJ, 630, 82
\bibitem[]{} P\'erez-Gonz\'alez, P.G. et al. 2005, ApJ, 630, 82
 
\bibitem[]{} Pettini, M., Steidel, C.C., Adelberger, K.L., Dickinson, M., Giavalisco, M. 2000, AJ, 528, 96

\bibitem[]{} Sanders, D.B., Soifer, B.T., Elias, J.H., Madore, B.F., Matthews, K., Neugebauer, G.; Scoville, N. Z. 1988, ApJ, 325, 74

\bibitem[]{} Schweizer, F. 1996, AJ, 111, 109
 
\bibitem[]{} Shapley, A.E., Steidel, C.C., Pettini, M., Adelberger, K.L.  2003, ApJ, 588, 65
 
\bibitem[]{} Steidel, C.C., Giavalisco, M, Pettini, M., Dickinson, M., Adelberger, K.L. 1996, ApJ, 462, L17
 
\bibitem[]{} Steidel, C.C., Shapley, A.E.,  Pettini, M., Adelberger, K.L., Erb, D.K., Reddy, N.A., Hunt, Matthew P  2004, ApJ, 604, 534

\bibitem[]{} Tadhunter, C.N., Fosbury, R.A.E., Quinn, P.J. 1989, MNRAS, 240, 225

\bibitem[]{} Tadhunter, C.,  Robinson, T.G., Gonz\' alez Delgado, R.M.,  Wills, K.,  Morganti, R. 2005, MNRAS, 356, 480

\bibitem[]{} Tadhunter, C.N., Dicken, D., Holt, J., et al. 2007, ApJ, 661, L13

\bibitem[]{} Tenorio-Tagle, G., Silich, S.A., Kunth, D., Terlevich, E, Terlevich, R.J. 1998, MNRAS, 309, 332

\bibitem[]{} Toorem, A.  \& Toorem, J. 1972, ApJ, 178, 623

\bibitem[]{} Tremonti, C., .A., Calzetti, D., Leitherer, C., Heckman, T.M. 2001, AJ, 555, 322

\bibitem[]{} Veilleux, S., Kim, D.-C.,  Sanders, D.B. 1999, ApJ, 522, 113
 
\bibitem[]{} Williams, R.E., Blacker, B.,  Dickinson, M. et al. 1996, ApJ, 112, 1335
 
 
 
 
\end{thebibliography}
\end{document}